\documentclass{fundam}

\publyear{25}
\papernumber{2}
\volume{194}
\issue{4}
\theDOI{10.46298/fi.14346}

\versionForARXIV

\usepackage{url}
\usepackage[ruled,lined]{algorithm2e}
\usepackage{graphicx}
\usepackage{tikz}
\usetikzlibrary{automata, positioning, arrows}
\usepackage[utf8]{inputenc}

\usepackage{amsmath}
\usepackage{amssymb}
\usepackage[T1]{fontenc}

\usepackage{hyperref}

\usepackage{enumitem}

\usepackage{pifont}%
\newcommand{\cmark}{\ding{51}}%
\newcommand{\xmark}{\ding{55}}%

\begin{document}

\title{Privacy for Quantum Annealing. Attack on Spin Reversal Transformations in the Case of Cryptanalysis}

\author{Mateusz Leśniak\\
Department of Cryptology \\
NASK National Research Institute, Kolska Str. 12, Warsaw, Poland\\
mateusz.lesniak{@}nask.pl
\and Michał Wroński\\
Department of Cryptology \\
NASK National Research Institute, Kolska Str. 12, Warsaw, Poland \\
michal.wronski{@}nask.pl} 

\maketitle

\runninghead{M.Leśniak, M.Wroński}{Privacy for Quantum Annealing}

\begin{abstract}
This paper demonstrates that applying spin reversal transformations, commonly known as a sufficient method for privacy enhancement in problems solved using quantum annealing, does not guarantee privacy for all possible cases. We show how to recover the original problem from the Ising problem obtained using spin reversal transformation when the resulting problem in Ising form represents the algebraic attack on the $E_0$ stream cipher. A small example illustrates how to retrieve the original problem from that transformed by spin reversal transformation. Moreover, we show that our method is efficient also for full-scale problems.
\end{abstract}


\section{Introduction} \label{sec: introduction}
    Quantum optimization is a highly complex process. Despite this, it is gaining considerable popularity \cite{Yarkoni:2022}. Due to its significant structural requirements, it is usually implemented as a cloud service. The delivered products allow us to solve a wide range of problems. In this paper, we focus on two specific aspects of quantum computation:
    \begin{itemize}[nosep]
        \item Quantum Annealing, introduced by Tadashi Kadowaki and Hidetoshi Nishimori in \cite{PhysRevE.58.5355};
        \item Quantum Approximate Optimization Algorithm, introduced by Edward Farhi et al. in \cite{farhi2014quantumapproximateoptimizationalgorithm}.
    \end{itemize}
    As presented in \cite{Yarkoni:2022, DWave:Applications}, among the application areas of annealing, we can distinguish traffic flow optimization, logistics, vehicle routing problems, finance, and quantum simulation. As interest in quantum optimization grows, so does the need for methods to keep the computations performed private. The classical homomorphic encryption approach \cite{Brakerski:2014, Gentry:2013, Rivest1978ONDB} does not apply to quantum optimization. In this case, dedicated methods, such as \cite{SpinReversalTransformations}, must be used. 

    In this paper, we focus only on the application method described in \cite{SpinReversalTransformations} and the confidentiality of the data used in the calculations. We attack the scheme shown in \cite{SpinReversalTransformations}. A pen-and-paper example of the attack supports the theoretical description of the attack.

    This paper is organized as follows. Section~\ref{sec: background} provides an overview of possible privacy-preserving methods, the assumed flow of communication, and a description of the models used. Section~\ref{sec: spin reversal transformation} describes the attack framework and the attack method's details. Section~\ref{sec: details of the proposed attack} introduces our attack, while Section~\ref{sec: a practical example of the proposed attack} demonstrates its practical feasibility using a pen-and-paper example. Finally, Section~\ref{sec: summary and future work} concludes the paper.

\section{Background} \label{sec: background}
     This section discusses privacy aspects in communication flow and the assumptions made about the adversary. It also presents the basics of optimization models and their transformations.
\subsection{Quantum cloud services and privacy}

\begin{figure}[ht]
        \centering
        \includegraphics[width=0.75\linewidth]{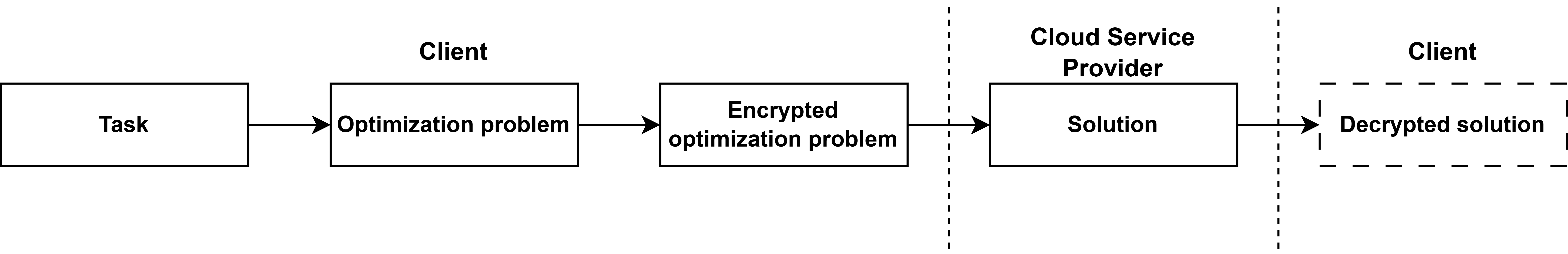}
        \caption{Data flow considered.}
        \label{fig: flow}
    \end{figure}

    Figure \ref{fig: flow} considers the communication flow. The client has a specific task that he wants to realize using optimization. In the first step, an optimization problem is formulated for the given task. The following optional step is to encrypt the resulting optimization problem. Assuming ideal communication conditions or having your own quantum computer, this step can be skipped as potentially unnecessary. However, in real-world conditions, failing to do this seriously threatens the privacy of the computations performed. After encryption, the problem is passed to a computer that can perform the indicated optimization. After optimization, a solution is returned. If the problem has been encrypted, it will be necessary to decrypt the solution received further.

    Due to the high cost of quantum infrastructure solutions \cite{Memon:2024}, quantum computing is typically implemented using remote access. Such a solution requires the transfer of data to an outsourcer. Due to the inability to verify the cloud service provider's intentions and the lack of control over the data flow on its side, the model discussed below may be assumed. 
    
    The adversary is between the client and the quantum cloud service. All communication goes through the adversary, which has the ability to perform any operation on the received problem. This paper assumes that adversaries eavesdrop on intercepted communications and pass them forward.
    
    With the growing number of cloud service providers, the threat is increasing. Service providers are in high demand, and their availability is severely limited. This leads to a situation where a new provider offering competitive services can gain popularity rapidly. The internal structure of cloud services remains a mystery, and it is unknown if there is no eavesdropping between the client and the actual computer. Such a new, untrusted provider may be malicious, and communications between the user and the quantum computer may be intercepted.

    In specific cases, such an arrangement is unacceptable. There are few critical applications of quantum optimization where the potential leak of private data involves serious consequences. Transferring the data to an external party is not permissible for such use. In the case of portfolio optimization, such a situation may lead to a financial benefit for the cloud service provider, which could use the optimal solution before passing it to the customer. Another branch of applications for which computational privacy is significant is cryptanalysis. As presented in \cite{Jiang2018}, it may be used in factorizing numbers, solving the discrete logarithm \cite{wronski2021index, LogarytmWronski, zolnierczyk2023searching, wronski2024transformation, wronski2024base} or \cite{burek2022algebraic,cryptoeprint:2023/1502} in algebraic attacks on symmetric ciphers. Computing without privacy in this situation could compromise classified or strategic data.

    Just as in the case of classical cloud computing, the solution to the problem is homomorphic encryption \cite{Rivest1978ONDB}; in the case of the discussed issue, the solution may be analogous to specialized privacy-preserving methods. Privacy in quantum computing is a promising area of research. Secure protection algorithms will expand the market for quantum services. As stated in \cite{ayanzadeh2023enigma}, Secure Quantum Computing can be divided into two groups of methods:
    \begin{itemize}[nosep]
        \item Blind Quantum Computing is a technique that allows outsourcing computations without disclosing the computation's details to the server. This group of protocols is often impractical and requires the customer to have quantum hardware or the participation of several servers.
        \item Quantum Homomorphic Encryption involves performing calculations on encrypted data. The result of the calculation, after decryption, is the result of the original problem.
    \end{itemize}
    This paper focuses on the spin reversal transformation method presented mainly in \cite{SpinReversalTransformations}. The authors propose homomorphic encryption for quantum annealing, which they believe protects the details of quantum annealing instances against a malicious cloud. As stated in the paper, the scheme runs on the Ising model, so it can also be used for privacy preservation in the Quantum Approximate Optimization Algorithm. 
        
\subsection{A description of the models used}
    As mentioned earlier, using the Quantum Approximate Optimization Algorithm or Quantum Annealing requires presenting the problem in a specific form. One such form is the Ising model, which has existed since 1920. Ernst Ising and Wilhelm Lenz introduced it as a description of magnetic materials. Despite its original application, the model has also been applied in combinatorial optimization. As mentioned in \cite{vertogen1973ising}, each Ising problem can be viewed as a minimizing expression presented as Equation \eqref{eq:ising}
    \begin{equation}
        \displaystyle
        \label{eq:ising}
        f(s) = \sum_{i=0}^{n-1} h_i s_i + \sum_{i,j=0}^{n-1} J_{i,j} s_i s_j .
    \end{equation}
    Vector \(s\) is called the state, and each variable \(s_i \in \{-1,1\}\) is called a spin. In practice, the following are used to characterize the problem: 
    \begin{itemize}[nosep]
        \item a vector of biases:
        \begin{equation}
            \displaystyle
            \label{eq:bias vector}
            h = \begin{bmatrix} h_0 & h_1 & \cdots & h_{n-1} \end{bmatrix}^T;
        \end{equation}
        \item a matrix describing connections between variables:
        \begin{equation}
            \displaystyle
            \label{eq:connection matrix}
            J = \begin{bmatrix} J_{0, 0} & J_{0, 1} & \cdots & J_{0, n-1} \\ J_{1, 0} & J_{1, 1} & \cdots & J_{1, n-1} \\ \vdots & \vdots & \ddots & \vdots \\ J_{n-1, 0} & J_{n-1, 1} & \cdots & J_{n-1, n-1} \end{bmatrix}.
        \end{equation}
    \end{itemize}
    In this case, communication with the quantum cloud requires sending vector of biases \(h\) and a connection matrix \(J\), presented as Equations \eqref{eq:bias vector} and \eqref{eq:connection matrix} respectively.

    An alternative model is the QUBO model, which uses binary variables \(x \in \{0,1 \}\). Quadratic unconstrained binary optimization can also be viewed as minimizing the specific expression shown as Equation \eqref{eq:qubo}
    \begin{equation}
        \displaystyle
        \label{eq:qubo}
        f(x) = \sum_{i=0}^{n-1} Q_{i,i} x_i + \sum_{i,j = 0}^n Q_{i,j} x_i x_j,
    \end{equation}
    where vector \(x\) contains \(n\) binary variables.

    As presented in \cite{LucasIsing}, there exist Ising formulations of many NP problems, such as graph partitioning \cite{Fu_86}, the knapsack problem \cite{Karp1972}, graph coloring \cite{Karp1972}, or the traveling salesman \cite{Karp1972} problem. However, it is easier for some tasks to formulate the problem in QUBO form. Such tasks include algebraic attacks on symmetric ciphers, as described in \cite{burek2022algebraic, burek2022speck, cryptoeprint:2023/1502}. 
    Applying privacy-preserving methods for such problems may require transitioning between the QUBO and Ising models and vice versa.

    There is a simple transformation between the Ising model and the QUBO problem. Its main idea is to perform the following substitutions to change spin variables into binary variables. When moving from the Ising model to the QUBO: 
    \begin{equation*}
        \displaystyle
        x_i = \frac{1}{2} \cdot (s_i + 1).
    \end{equation*}
    When moving from the QUBO to the Ising model: 
    \begin{equation*}
        \displaystyle
        s_i = 2 x_i - 1.
    \end{equation*}
    As described in \cite{ayanzadeh2023enigma}, coefficients of each matrix can be determined using matrix coefficients for the corresponding problem. When moving from the Ising model to the QUBO problem, the connection matrix is determined as follows:
    \begin{equation}
        \displaystyle
        \label{eq:ising to qubo:matrix}
        \begin{array}{c}
            Q_{i,j} = 4 J_{i,j}, \\
            Q_{i,i} = 2\bigl(h_i - \sum_j J_{i,j} - \sum_j J_{j,i}\bigr).
        \end{array}
    \end{equation}
    At the transition in the opposite direction, the bias vector is determined as follows:
    \begin{equation}
        \displaystyle
        \label{eq:qubo to ising:bias}
        h_{i} = \frac{Q_{i,i}}{2} + \frac{\sum_j Q_{i,j} + \sum_j Q_{j,i}}{4}.
    \end{equation}
    The connection matrix is determined as follows:
    \begin{equation}
        \displaystyle
        \label{eq:qubo to ising:connection}
        J_{i,j} = \frac{Q_{i,j}}{4}.
    \end{equation}
    In the rest of the paper, we skip transitions between models, considering this operation trivial.
        
\section{Spin Reversal Transformation in detail} \label{sec: spin reversal transformation}
    This section briefly describes the application of spin reversal transformation to preserve privacy. This idea is mainly presented in \cite{SpinReversalTransformations} and partly in \cite{ayanzadeh2023enigma}. The method utilizes a random sequence to reverse a given sign in an Ising problem instance. According to the authors, an adversary, having intercepted a concealed problem, cannot reconstruct the original problem without knowing the original key. 

\subsection{Description of the algorithm}
    The encryption scheme described in \cite{SpinReversalTransformations} is based on spin reversal transformation, also called a gauge transformation \cite{Gauge2019}. The transformation uses a binary string \(x\) to change the signs of the selected coefficients. After applying the mentioned transformation, Equation \eqref{eq:ising} is transformed as follows:
    \begin{equation}
        \displaystyle
        f^{*}(s^*) = \sum_{i=0}^{n-1} (-1)^{x_i} h_i s^*_i + \sum_{i,j=0}^{n-1} (-1)^{x_i + x_j} J_{i,j} s^*_i s^*_j.
    \end{equation}
    The same sequence \(x\) must be used to determine the original solution. The corresponding solution can be determined according to Equation \ref{eq:solution}:
    \begin{equation}
        \displaystyle
        \label{eq:solution}
        s_i = (-1)^{x_i} s_i^{*}.
    \end{equation}
    It is important to note that the minimal energy of the instance does not change. As shown in \cite{SpinReversalTransformations} and \cite{Gauge2019}, the solution to the concealed problem, when uncovered, is the solution to the original problem.

    The described transformation can be outlined as a simple scheme, as in \cite{SpinReversalTransformations}:
    \begin{enumerate}[nosep]
        \item The client generates a secret key \(x = (x_0, x_1, \ldots, x_{n-1})\);
        \item The client computes \(h^*\) and matrix \(J^*\):
        \begin{eqnarray}
                h_i^* = (-1)^{x_i} h_i,\\
                J_{i,j}^* = (-1)^{x_i + x_j} J_{i,j},
        \end{eqnarray}
        \item The client sends the concealed problem to the quantum cloud service and receives the solution~\(s^*\);
        \item The client retrieves the solution using Equation \eqref{eq:solution}. 
    \end{enumerate}
        
\section{Details of the proposed attack} \label{sec: details of the proposed attack}
    This paper argues that the privacy-enhancing mechanism presented in \cite{SpinReversalTransformations} fails in many instances. As a counterexample, we demonstrate that the method fails when the technique of transforming the stream cipher cryptanalytic problem is known. Specifically, we show that, without prior knowledge of the random sequence used to conceal the given Ising problem for cryptanalysis of the \(E_0\) cipher, one can easily retrieve the original problem using the intercepted data. Furthermore, by knowing the solution to the concealed problem, one can also deduce the solution to the original problem. It also allows the recovery of the concealment key and the data on which the client wants confidentiality (the key hidden in the solution to the problem sent, ciphertexts, or keystream). With the concealment key, the adversary can recover communications encrypted with this key. The design of our attack involves three phases:
    \begin{enumerate}[nosep]
        \item \textbf{Parameterization (optional)}: Based on knowledge of the type of optimization task and access to the oracle \(\phi\), a parameterized matrix is constructed. It can be implemented before the attack and only once for a given optimization task. Subsequent attacks use the predetermined matrix.
        \item \textbf{System setup}: Using the parameterized matrix and the intercepted encrypted optimization problem \(h_{i}^*, J_{i,j}^*\), a system of linear equations is created. 
        \item \textbf{Solving}: The obtained system of equations is solved, determining the keystream used to create the optimization task and the concealment key. 
    \end{enumerate}
    
\subsection{The problem for the \(E_0\) cipher}
    We present an attack on the mentioned scheme using the QUBO problem corresponding to the cryptanalysis of the $E_0$ cipher, presented in \cite{LesniakE0}. As the authors point out, the reduction shown in that paper in the next few years can be practically embedded in a commercially available quantum annealer. This opens up the possibility of application to real cryptanalysis, for which it is essential to maintain the privacy of transmitted data and recovered keys. 
    
    Our proposed attack can also be applied to other highly structured optimization problems. The following example was chosen for its potential practicality and its scalability.
    
    Below is a brief overview of the construction of the $E_0$ cipher. A full description of the cipher can be found in \cite{Bluetooth53}. The cipher is built from three elements:
    \begin{itemize}[nosep]
        \item Four shift registers with linear feedback, specified by the following primitive polynomials \(f_i(x)\):
        \begin{equation*}
            \displaystyle
            \begin{array}{cc}
                 L_1: & f_1(x) = x^{25} \oplus x^{20} \oplus x^{12} \oplus x^{8} \oplus 1, \\
                 L_2: & f_2(x) = x^{31} \oplus x^{24} \oplus x^{16} \oplus x^{12} \oplus 1, \\
                 L_3: & f_3(x) = x^{33} \oplus x^{28} \oplus x^{24} \oplus x^{4} \oplus 1, \\
                 L_4: & f_4(x) = x^{39} \oplus x^{36} \oplus x^{28} \oplus x^{4} \oplus 1. 
            \end{array}
        \end{equation*}
        \item Summation Combiner Logic, computing the two-bit value \(s_{t+1}\):
        \begin{equation*}
            \displaystyle
            F_1: s_{t+1} = \Bigl \lfloor \frac{\sum_{i=1}^{4}x_i + c_t}{2} \Bigr \rfloor .
        \end{equation*}
        \item Blend Register, calculating the two-bit value of \(c_{t+1}\) using the bijections described in the cipher specification:
        \begin{equation*}
            \displaystyle
            F_2: c_{t+1} = s_{t+1} \oplus T_1[c_t] \oplus T_2[c_{t-1}],
        \end{equation*}
        where \(T_i\) are linear mappings:
        \begin{equation*}
            \begin{array}{cl}
                 T_1: &  (x_1, x_0) \to (x_1, x_0), \\
                 T_2: &  (x_1, x_0) \to (x_0, x_1 \oplus x_0).
            \end{array}
        \end{equation*}
        \item Each keystream bit is determined by the value from shift registers with linear feedback, \(x_0\), \(x_1\), \(x_2\), \(x_3\), and one bit from the Blend Register:
        \begin{equation*}
            \displaystyle
            z_i = x_1 \oplus x_2 \oplus x_3 \oplus x_4 \oplus c_t^0.
        \end{equation*}
    \end{itemize}
    Each bit of the keystream is described by eight equations: 
    \begin{equation}
        \displaystyle
        \label{eq:equations}
        \begin{cases}
            f_0: z_t = l_{t+1} \oplus m_{t+7} \oplus n_{t+1} \oplus o_{t+7} \oplus c^0_t, \\
            f_1: l_{t+25} = l_{t} \oplus l_{t+5} \oplus l_{t+13} \oplus l_{t+17}, \\
            f_2: m_{t+31} = m_{t} \oplus m_{t+7} \oplus m_{t+15} \oplus m_{t+19}, \\
            f_3: n_{t+33} = n_{t} \oplus n_{t+5} \oplus n_{t+9} \oplus n_{t+29}, \\
            f_4: o_{t+39} = o_{t} \oplus o_{t+3} \oplus o_{t+11} \oplus o_{t+35}, \\
            f_5: c^1_{t+1} = s^1_{t+1} \oplus c^1_{t} \oplus c^0_{t-1}, \\
            f_6: c^0_{t+1} = s^0_{t+1} \oplus c^0_{t} \oplus c^1_{t-1} \oplus c^0_{t-1}, \\
            f_7: 4 s^1_{t+1} + 2 s^0_{t+1} + \beta = l_{t+1} + m_{t+7} + n_{t+1} + o_{t+7} + 2 c^1_t + c^0_t.
        \end{cases}
    \end{equation}
    
\subsection{From algebraic attack to quantum optimization} \label{sec: from algebraic attack to quantum optimization}
    This section briefly describes the transformation of the algebraic attack to the QUBO optimization problem. The main idea was presented in \cite{burek2022algebraic} and adapted to \(E_0\) in \cite{LesniakE0}. In the first step, a system of equations is generated. To recover the initial internal state of the cipher, we need 128 bits of the keystream. As mentioned, each stream bit is described by eight equations. In total, the system describing the cipher will have 1024 equations. Below, we show how to perform transformations of the resulting system in a few steps:
    \begin{enumerate}[nosep]
        \item Equations \(f_i, i = \overline{0,6}\) are transformed into equations with binary variables and integer coefficients. Equation \(f_7\) does not require such a transformation. From the construction of the cipher, it is already in this form:
        \begin{equation*}
            \displaystyle
            f_i' \equiv 0 (\text{mod }2) \to f_i - 2k_i = 0.
        \end{equation*}
        Each integer variable \(k_i\) is bounded, \(k_i \leq \lfloor \frac{f_i^\text{max}}{2} \rfloor\), where \(f_i^\text{max}\) is the maximum value of the selected polynomial.
        \item In the standard transformation, the equations must be linearized in the next step. However, this step is skipped in this paper; the \(E_0\) design induces no nonlinear equations.
        \item Then variables \(k_i\) are replaced with binary variables. Each \(k_i\) requires \(bl(f_i^\text{max})\) new binary variables, where \(bl(x)\) denotes the bit-length of \(x\).
        \item In the final step, polynomial \(F'_\text{Pen}\) is determined according to Equation \eqref{eq:sum}:
        \begin{equation}
            \displaystyle
            \label{eq:sum}
            F'_\text{Pen} = \sum_{i=0}^{m-1} (f'_i)^2,
        \end{equation}
        The standard components with a penalty are omitted because they are zero. Finally, the constant present in the polynomial is subtracted from the resulting polynomial. 
    \end{enumerate}
    The final matrix obtained has a size of \(N = 2728\) variables and \(20598\) non-zero coefficients, which is \(0.55\%\) of all matrix elements.

\subsection{How to identify the problem and parameterize it?} \label{sec:How to identify the problem and parameterize it?}
    Quantum service providers require data to identify the user. Among such data, we can distinguish:
    \begin{itemize}[nosep]
        \item first and last name;
        \item email address;
        \item company name, job title, field of study;
        \item the purpose of computer access.
    \end{itemize}
    Using them, an untrusted provider can identify the client and the research area in which the client works. Identifying the area of study will allow the data obtained to be matched with one of the known problems. Problems vary significantly in size, density, and coefficient range. We assume that resources cannot freely change the connectivity (density) or size of the problem. 

    The lowest layer and critical element of the attack is the parameterization of the matrix of the selected optimization problem. The parameterization details will depend on the type of optimization task. However, the general principle remains the same. Below is an idea of how it can be realized for an algebraic attack on an \(E_0\) cipher.

    As described earlier, the selected optimization problem depends on the given keystream. To parameterize its matrix, we need to identify the coefficients of the matrix that depend on the \(z_i\). As shown in Equation \eqref{eq:equations}, only one of the equations depends on the bits of the keystream. We highlight two methods: the equation analysis dependent on \(z_i\) and the algorithmic approach based on the methods of construction of the optimization problem, where an oracle \(\phi\) is created. 
    
    The first method requires analyzing the equations. According to the method of transforming the algebraic attack described earlier, we focus on Equation \eqref{eq:sum}. For each bit of the keystream, there will be a component in the final polynomial shown in:
    \begin{equation*}
        \displaystyle
        z_t + l_{t+1} + m_{t+7} + n_{t+1} + o_{t+7} + c^0_t - 2K = 0 \Big /^2 .
    \end{equation*}
    After squaring, we get six coefficients depending on the keystream bits: \(2z_t \cdot l_{t+1}, 2z_t \cdot m_{t+7}, 2z_t \cdot n_{t+1}, 2z_t \cdot o_{t+7}, 2z_t \cdot c_{t}^{0}, -4z_t \cdot K\). Note that \(K\) is an integer variable and should be replaced with a binary variable. The coefficients resulting from the squaring of other equations are constant. With the above knowledge and using Equation \eqref{eq:qubo to ising:bias}, equations dependent on the keystream occurring in the vector of biases can be computed.
    
    The alternative method does not require direct analysis of equations. It is more generic and can be used universally for any problem for which the construction of the optimization problem is known. The coefficients of the matrix depend linearly on secret information. This is the situation for the \(E_0\) cipher and most stream ciphers. Assume that an oracle \(\phi\) is given which, for any keystream, will return an Ising model, as in Equation \eqref{eq:oracle}. The oracle can be constructed based on known publications on selected optimization problems, similar to the earlier description in Section \ref{sec: from algebraic attack to quantum optimization}. The chosen method requires \(|z| + 1\) queries to create a parameterized matrix for the keystream of length \(|z|\). 
    
    To determine the coefficients depending on a specific bit of the keystream, we use all streams with a Hamming weight of 1. Such a stream can be denoted as \(s_i\) and is presented as Equation \eqref{eq:stream}:
    \vspace*{-1ex}
    \begin{equation}
        \displaystyle
        \label{eq:stream}
        s_i = \{\underbrace{0, \ldots, 0}_{i-1}, \underset{i}{1}, \underbrace{0, \ldots, 0}_{n-i}\}.
    \end{equation}
    We denote the Ising models corresponding to the streams \(s_i\) as \(h_i\) and \(J_i\). They are determined using the oracle \(\phi\), as shown in Equation \eqref{eq:oracle}:
        \vspace*{-2ex}
    \begin{equation}
        \displaystyle
        \label{eq:oracle}
        h_i, J_i \gets \phi(s_i).
    \end{equation}
    Additionally, we denote \(h_{\infty}\) and \(J_{\infty}\) as the Ising model for a stream with Hamming weight equal to zero. Then, the parameterized problem is computed as follows:
        \vspace*{-1ex}
    \begin{equation}
        \displaystyle
        \label{eq:paraemtrized}
        \begin{array}{c}
            h^P = h_{\infty} + \sum_{i=0}^n (h_i - h_{\infty}) \cdot z_i,\\
            J^P = J_{\infty} + \sum_{i=0}^n (J_i - J_{\infty}) \cdot z_i,
        \end{array}
    \end{equation}
    where \(z_i\) is a variable. 
    \newline
    The idea of this parameterization method is illustrated in the example in Section \ref{sec: example}.

\subsection{Executing the attack, the most straightforward phase} \label{sec: executing the attack}
    An attack is performed using a parameterized problem. A vector of biases is sufficient to perform the attack. 
    
    First, a system of linear equations is constructed. The encrypted vector of biases \(h^*\) is juxtaposed with a parameterized one \(h^P\) to form a system of equations:
        \vspace*{-1ex}
    \begin{equation*}
        h^* = h^P = \begin{bmatrix}
            h^*_0 \\ h^*_1 \\ \vdots \\ h^*_N
        \end{bmatrix} = \begin{bmatrix}
            h^P_0 \\ h^P_1 \\ \vdots \\ h^P_N
        \end{bmatrix}.
    \end{equation*}
        \vspace*{-2ex} \\
    As the design of the parameterized matrix shows, each element of \(h^P\) can be presented as a sum of rational number \(\texttt{const}_i\) and a linear combination of selected bits of the keystream. We denote the set of indices of the relevant bits for element \(i\) as \(\mathcal{A}_i\). The equations formed in the proposed way can be represented by Equation \eqref{eq:attack}:
    \begin{equation}
        \displaystyle
        \label{eq:attack}
        h_i^* = \sum_{j \in \mathcal{A}_i} z_j + \texttt{const}_i,
    \end{equation}
        \vspace*{-3ex} \\
    where \(\texttt{const}_i \in \mathbb{Q}\).

    The next step is to analyze each resulting equation in the correct order. According to how the problem is constructed, there will be more equations than unknowns in the created matrix. From among all the equations, we choose some set of sufficient size and reduce the equations to the form presented in Equation \eqref{eq:simplify}:
    \begin{equation}
        \displaystyle
        \label{eq:simplify}
        z_k + b_i = h_i^*,
    \end{equation}
    where \(b_i \in \mathbb{Q}\) is sum of \(\texttt{const}_i \) and known keystream bits. As \(k\) we denotes the index of the keystream variable occurring in this equation.

    If the transformation of all the equations is impossible, we rearrange the system to an upper triangular form. Then, an equation analysis is performed, starting with the last equation. In the next steps, successive equations will be reduced, considering the previous solutions and performing an analysis.

    As can be seen from the definition of the optimization problem under consideration, the variable \(z_k\) is a binary variable, \(z_k \in \{0, 1\}\). Therefore, the relation \eqref{eq:relation} should be satisfied:
    \begin{equation}
        \displaystyle
        \label{eq:relation}
        h_i^* - b_i \in \{0, 1\}.
    \end{equation}
    Due to the use of encryption, two situations can occur: the condition will be met or not. If the above relation is satisfied, then:
    \begin{itemize}[nosep]
        \item the relevant coefficient has not been concealed;
        \item the designated bit of the concealment key is \(x_i = 0\);
        \item the designated bit of the keystream is \(z_k = h_i^* - b_i\).
    \end{itemize}
    Otherwise, the given coefficient is concealed and:
    \begin{itemize}[nosep]
        \item the designated bit of the concealment key is \(x_i = 1\);
        \item the designated bit of the keystream is \(z_k = - h_i^* - b_i\).
    \end{itemize}
    In most cases, solving the equation above gives the bit of the keystream \(z_k\) with a probability of \(1\). If one finds that \(z_k \in \{0, 1\}\) regardless of whether a coefficient is concealed, one can check if other equations where \(z_k\) appears are correct. 
    
    The above procedure allows the recovery of the used keystream. The remaining bits of the concealment key can be determined by comparing the remaining coefficients. If the coefficients are opposite, the given coefficient has been concealed, and the corresponding concealment key bit is 1. For every bit of the keystream, retrieving the correct value of the bit requires only solving an affine equation, and the entire attack is swift. In addition, the attack allows the recovery of the concealment key. With the use of this key:
    \begin{itemize}[nosep]
        \item any further problem encrypted with the same key can be exposed;
        \item the solution to the problem sent may be unveiled, including knowledge of the secret key that the client wanted to recover.
    \end{itemize}
    
\section{A practical example of the proposed attack} \label{sec: a practical example of the proposed attack}
    To illustrate the correctness of the attack, an attack performed in practice is presented. A scaled-down \(E_0\) cipher was used to show the step-by-step operation of the attack. In addition, the execution of the attack on the full version is discussed.

\subsection{Example parameterization} \label{sec: example}
    In this section, we present a pen-and-paper example of parameterization. We are given an oracle \(\phi\) that returns bias vectors for given streams generated by a hypothetical stream cipher. Here, we focus on supporting the idea of parameterization with an example, so we omit the details of the hypothetical cipher. Assume that we are looking for a parameterized matrix corresponding to an algebraic attack using a 3-bit keystream \((z_0, z_1, z_2)\). We now follow the description presented earlier:
    \begin{enumerate}[nosep]
        \item We determine the bias vector for a stream with a Hamming weight of 0:
            \vspace*{-1ex}
        \begin{equation*}
            h_{\infty} = \phi(0,0,0) = \begin{bmatrix} 2 \\ 4 \\ 1 \end{bmatrix}.
        \end{equation*}
        \item We determine the bias vector for all streams with a Hamming weight of 1:
        \begin{equation*}
            \begin{array}{ccc}
                h_0 = \phi(1,0,0) = \begin{bmatrix} 3 \\ 5 \\ 2 \end{bmatrix}, &
                h_1 = \phi(0,1,0) = \begin{bmatrix} 2 \\ 5 \\ 1 \end{bmatrix}, &
                h_2 = \phi(0,0,1) = \begin{bmatrix} 2 \\ 4 \\ 7 \end{bmatrix}.
            \end{array}
        \end{equation*}
        \item We determine the differences \(h_i - h_{\infty}\), where \(i \in \{0, 1, 2\}\):
        \begin{equation*}
            \begin{array}{ccc}
                h_0 - h_{\infty} = \begin{bmatrix} 1 \\ 1 \\ 1 \end{bmatrix}, &
                h_1 - h_{\infty} = \begin{bmatrix} 0 \\ 1 \\ 0 \end{bmatrix}, &
                h_2 - h_{\infty} = \begin{bmatrix} 0 \\ 0 \\ 6 \end{bmatrix}.
            \end{array}
        \end{equation*}
        \item We determine the parameterized vector as:
        \begin{equation*}
            h^P = h_{\infty} + (h_0 - h_{\infty}) \cdot \texttt{z}_\texttt{0} + (h_1 - h_{\infty}) \cdot \texttt{z}_\texttt{1} + (h_2 - h_{\infty}) \cdot \texttt{z}_\texttt{2},
        \end{equation*}
            \vspace*{-1ex}
        so:
        \begin{equation*}
            h^P = \begin{bmatrix} 2 \\ 4 \\ 1 \end{bmatrix}
            +
            \begin{bmatrix} \texttt{z}_\texttt{0} \\ \texttt{z}_\texttt{0} \\ \texttt{z}_\texttt{0} \end{bmatrix} 
            + 
            \begin{bmatrix} 0 \\ \texttt{z}_\texttt{1} \\ 0 \end{bmatrix}
            +
            \begin{bmatrix} 0 \\ 0 \\ 6 \texttt{z}_\texttt{2} \end{bmatrix}
            = 
            \begin{bmatrix}
                2 + \texttt{z}_\texttt{0} \\
                4 + \texttt{z}_\texttt{0} + \texttt{z}_\texttt{1} \\
                1 + \texttt{z}_\texttt{0} + 6 \texttt{z}_\texttt{2} 
            \end{bmatrix}.
        \end{equation*}
    \end{enumerate}
    The searched parameterized matrix of the assumed algebraic attack was thus determined using four oracle queries. 
    
\subsection{An illustrative attack on a scaled-down instance}
    An instance of the cipher using LFSRs described by the following primitive polynomials was used for the attack:
    \begin{itemize}[nosep]
        \item \(L_1: f_1(x) = x^3 + x + 1,\)
        \item \(L_2: f_2(x) = x^3 + x + 1,\)
        \item \(L_3: f_3(x) = x^3 + x + 1,\)
        \item \(L_4: f_4(x) = x^3 + x + 1.\)
    \end{itemize}
    To simplify the generated problem as much as possible, the registers of the scaled cipher should be of equal length. The set of primitive polynomials of degree 3 is limited. For this reason, the selected polynomials are equal. The impact of such a solution on the cipher's security is not the subject of this paper. Such a configuration allows a pictorial representation of the proposed attack. 
    
    For the selected cipher instance, the following keystream was used:
    \begin{equation}
        \displaystyle
        \label{eq:keystream}
        z = (0, 0, 1, 0, 1, 1, 1, 1, 0, 0, 1, 0).
    \end{equation}
    For the given sequence, 96 equations were generated. Based on these equations, a QUBO problem with 240 variables was determined. The problem was transformed into an Ising model and encrypted using the idea presented in \cite{SpinReversalTransformations}, as described in Section \ref{sec: spin reversal transformation}. Due to the size of the matrix, we cannot include the entire key here. Below, as Equation \eqref{eq:concealment key}, we present selected key bits corresponding to the coefficients of the vector at the positions analyzed:
    \begin{equation}
        \displaystyle
        \label{eq:concealment key}
        x = (\underset{121}{1},
        \underset{131}{1},
        \underset{141}{1},
        \underset{151}{0},
        \underset{161}{1},
        \underset{171}{1},
        \underset{181}{0},
        \underset{191}{1},
        \underset{201}{1},
        \underset{211}{0},
        \underset{221}{0},
        \underset{231}{1}).
    \end{equation}
    Values under consecutive bits indicate the position number of the specified key bit.

    An oracle was built based on \cite{LesniakE0}. Using it, a parameterized problem was generated according to the idea in Section \ref{sec:How to identify the problem and parameterize it?}. To obtain the entire parameterized bias vector, 13 queries were performed. The equations presented as Equation \eqref{eq:example} are selected to recover the keystream from this vector. As described in Section \ref{sec: executing the attack}, such a set of equations was chosen so that it would be possible to analyze them one by one and determine the entire keystream. Out of 240 equations in a parameterized matrix, 12 were selected for further analysis:
    \begin{equation}
        \displaystyle
        \label{eq:example}
        \begin{cases}
            f_{121}: -z_0 - 1 = 1, \\
            f_{131}: -z_1 - 1 = 1, \\
            f_{141}: -z_2 - 1 = 2, \\
            f_{151}: -z_3 - 1 = -1, \\
            f_{161}: -z_4 - 1 = 2, \\
            f_{171}: -z_5 - 1 = 2, \\
            f_{181}: -z_6 - 1 = -2, \\
            f_{191}: -z_7 - 1 = 2, \\
            f_{201}: -z_8 - 1 = 1, \\
            f_{211}: -z_9 - 1 = -1, \\
            f_{221}: -z_{10} - 1 = -2, \\
            f_{231}: -z_{11} - 1 = 1. 
        \end{cases}
    \end{equation}
    The equation analysis is shown in Table \ref{tab:analysis}. Based on it, the coefficients with changed signs were determined. 
    \begin{table}[ht]
        \centering
        \caption{Analysis of selected equations. Determination of part of the key bits.}
        \label{tab:analysis}
        \begin{tabular}{|c|c|c|c|} \hline
            \(i\) & Equation  & \(z_i \in \{0, 1\}\) & \(x_i\) \\ \hline
            121 & \(-z_0 - 1 = 1 \to z_0 = -2\) & \xmark & 1\\ \hline
            131 & \(-z_1 - 1 = 1 \to z_1 = -2\) & \xmark & 1\\ \hline
            141 & \(-z_2 - 1 = 2 \to z_2 = -3\) & \xmark & 1\\ \hline
            151 & \(-z_3 - 1 = -1 \to z_3 = 0\) & \cmark & 0\\ \hline
            161 & \(-z_4 - 1 = 2 \to z_4 = -3\) & \xmark & 1\\ \hline
            171 & \(-z_5 - 1 = 2 \to z_5 = -3\) & \xmark & 1\\ \hline
            181 & \(-z_6 - 1 = -2 \to z_6 = 1\) & \cmark & 0\\ \hline
            191 & \(-z_7 - 1 = 2 \to z_7 = -3\) & \xmark & 1\\ \hline
            201 & \(-z_8 - 1 = 1 \to z_8 = -2\) & \xmark & 1\\ \hline
            211 & \(-z_9 - 1 = -1 \to z_9 = 0\) & \cmark & 0\\ \hline
            221 & \(-z_{10} - 1 = -2 \to z_{10} = 1\) & \cmark & 0\\ \hline
            231 & \(-z_{11} - 1 = 1 \to z_{11} = -2\) & \xmark & 1\\ \hline
        \end{tabular}
    \end{table}
    Finally, based on the above analysis, the keystream was determined. 

\begin{figure}[ht]
        \centering
        \includegraphics[width=0.95\textwidth]{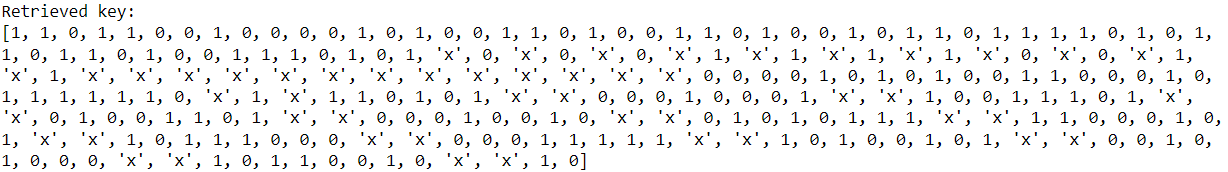}
        \caption{Key recovered as a result of the experiment.}
        \label{fig:retrived key}
    \end{figure}

    The relevant part of the concealment key was determined using the remaining elements of the vector. The recovered key, the result of the developed script implementing the attack, is shown in Figure \ref{fig:retrived key}. Missing bits, denoted as \(x\) in Figure \ref{fig:retrived key}, of the key do not affect the ciphertext. It is impossible to change the sign of a coefficient equal to 0. 

    We can verify the correctness in two ways:
    \begin{itemize}[nosep]
        \item performing a comparison of the sequence shown in Equation \eqref{eq:concealment key} with the last column of Table~\ref{tab:analysis};
        \item using the recovered keystream, the full bias vector is determined. Using the recovered key, the concealed vector is decrypted. Then, a comparison of both vectors is performed. 
    \end{itemize}
    Both of these methods confirm the correctness of the attack performed. 
        
\subsection{Attack on the full version of the \(E_0\) cipher}
    An attack on an encrypted problem corresponding to the cryptanalysis of the full \(E_0\) cipher can be performed similarly. Using a script in the SageMath environment:
    \begin{enumerate}[nosep]
        \item A problem corresponding to an algebraic attack on the \(E_0\) cipher is generated for the selected keystream. The problem has 2728 variables, and the used keystream has 128 bits.
        \item The resulting optimization problem is encrypted using Spin Reversal Transformation with a randomly generated key.
        \item A parameterized matrix is determined. It requires 129 oracle calls. The time needed to generate the parameterized matrix can be estimated at about 13.5 hours. 
        \item Based on the determined matrix, a system of 128 linear equations is arranged. 
        \item In the last step, the designated equations are analyzed one by one, and the key used and the stream for which the problem was generated is determined.
        \item Finally, the results obtained are verified. The recovered keystream is compared with the original one. The matrix is decrypted and compared with the problem generated for the recovered keystream.
    \end{enumerate}
    As in the rescaled example, the longest step is the parameterization of the matrix. However, the duration of the attack still allows it to be performed in practice.
    
    It should be noted that the script used is not optimized in any way, and the experiment only illustrates the disparity between the time of the actual phase of the attack and the time of the pre-computations.  
        
\section{Summary and future work} \label{sec: summary and future work}
    This paper shows a practical attack on a proposed scheme to ensure the privacy of problems sent to quantum computing clouds. In addition, a practical attack on a miniature version of the cipher was demonstrated to show the correctness of the attack. It should be noted that the work does not address the correctness of the scheme under attack. The proposed attack was prepared with assumptions in the original work and additional ones resulting from the functioning of available services.

    Further work should look for a secure way to protect cloud computing based on the Ising model. As an alternative to using an untrusted provider, methods can be developed that allow local problem solving using private infrastructure.
    
\bibliographystyle{fundam}
\bibliography{citations}

\end{document}